\documentclass[aps,prd,superscriptaddress,floatfix,nofootinbib,notitlepage,12pt]{revtex4-2}
\usepackage{aas_macros}
\usepackage{graphicx}
\usepackage{multirow}
\usepackage{amsmath,amssymb,amsfonts}
\usepackage{verbatim}
\usepackage{bm}
\usepackage{color}
\usepackage{ulem}
\usepackage{mathtools}
\usepackage{colortbl}
\usepackage[dvipsnames]{xcolor}
\usepackage{slashed}
\usepackage{braket}
\usepackage{adjustbox}

\usepackage[colorlinks]{hyperref}
\hypersetup{
    colorlinks=true,
    linkcolor=blue,
    filecolor=blue,      
    urlcolor=blue,
    citecolor=magenta,
    pdfpagemode=FullScreen
}

\newcommand{\gev}{\text{GeV}}

\def\mugaga{\mu_{\gamma\gamma}}
\def\mugaZ{\mu_{\gamma Z}}

\date{\today}

\begin{document}
\title{Revisiting Inert Doublet Model Parameters}

\author{Hamza Abouabid}
\email{hamza.abouabid@gmail.com}
\affiliation{D\'{e}partement de Math\'{e}matiques,
Facult\'{e} des Sciences et Techniques,
Universit\'{e} Abdelmalek Essaadi, B. 416, Tangier, Morocco.}

\author{Abdesslam Arhrib}
\email{aarhrib@gmail.com}
\affiliation{D\'{e}partement de Math\'{e}matiques,
Facult\'{e} des Sciences et Techniques,
Universit\'{e} Abdelmalek Essaadi, B. 416, Tangier, Morocco.}

\author{Ayoub Hmissou}
\email{ayoub1hmissou@gmail.com}
\affiliation{Laboratory of Theoretical and High Energy Physics, Faculty of Science, Ibnou Zohr University, B.P 8106, Agadir, Morocco.}

\author{Larbi Rahili}
\email{l.rahili@uiz.ac.ma/rahililarbi@gmail.com}
\affiliation{Laboratory of Theoretical and High Energy Physics, Faculty of Science, Ibnou Zohr University, B.P 8106, Agadir, Morocco.}

\begin{abstract}
    In this study, we aim  to see how much the actual measurement of the Z+photon and di-photon signal strength, $\mugaga$ and $\mugaZ$, could influence the allowed parameter space of the Inert Doublet Model (IDM), and to what extent such measurement can be aligned with the latest bound from XENON1T experiment on the spin-independent dark-matter-nucleon scattering cross-section. Also, by considering the new embedded scalars in the IDM (i.e., $H$, $A$ and $H^\pm$), a wide investigation of the one-loop radiative corrections to the trilinear Higgs coupling $hhh$ has been made in the light of the previous measurements

\end{abstract}

\maketitle

\section{Introduction}
\label{sec:introduction}

Following the discovery of the Higgs boson at the Large Hadron Collider (LHC) ~\cite{Chatrchyan:2012xdj,Aad:2012tfa}, the standard model (SM) of particle physics was crowned with great success and highly accurate predictions to date.
The LHC program has already performed so far several precise measurements of the Higgs coupling to SM particles.
These measurements demonstrate that the SM works well in explaining these observed phenomena at the electroweak scale.
Moreover, SM has no answer to a few problems like non-zero neutrino masses, dark matter (DM) and nongravitational interactions, etc. This has prompted the search for new physics (NP) beyond the Standard Model (BSM) by extending the SM to contain extra: real or complex singlets, doublets, triplets higgs fields.

The Inert Doublet Model (IDM)  ~\cite{Deshpande:1977rw,Cao:2007rm,Barbieri:2006dq} constitutes a simple and  a phenomenologically interesting extension of the SM Higgs sector which features a DM candidate. It is a version of
the 2 Higgs Doublet Model (2HDM) with an exact $\mathbb{Z}_2$ symmetry, consisting of adding an inert scalar doublet $H_2$ to the SM Higgs doublet $H_1$. The doublet $H_2$ is odd under the new discrete $\mathbb{Z}_2$ symmetry and does not couple with fermions,  does not develop a vacuum expectation value (VEV). Such modified version of the 2HDM with an exact  $\mathbb{Z}_2$  symmetry is motivated by having a potential source of a weakly interacting massive particle (WIMP)
as well as a possible explanation of the observed excess of cosmic-ray positrons \cite{LopezHonorez:2010eeh,Dolle:2009fn}.

The purpose of this paper is to investigate the effect of the current measurement  of the di-photon ($\mugaga$)   and the recent  Z-photon ( $\mugaZ$ ) signal strengths on the IDM parameter space. Since these 2 observables are sensitive only to
the charged scalar contributions, one can use them to set a constraint on the coupling $hH^+H^-$ once the charged scalar boson mass is fixed.  In the numerical scan, we will take into consideration all theoretical constraints
on the scalar sector of the model as well as LHC experimental measurement such $\mu_{\gamma\gamma}^{exp}$
and $\mu_{Z\gamma}^{exp}$ and the invisible decay of the SM higgs.
In addition, an update on  the effects of the extra scalars of the IDM
onto the radiative corrections to the triple Higgs coupling $hhh$ at the one-loop level is presented.
We also examine to what extent such measurements can be consistent with the latest bound from XENON1T
experiment on the spin-independent dark-matter-nucleon scattering cross-section.

The paper is structured as follows. In Sec.~\ref{sec:idm-model} we review the details of the  IDM including its scalar potential and the corresponding constraints. In section ~\ref{sec:signal-strength}, a detailed look at the signal strength $h\to \gamma \gamma$ and $h\to Z\gamma$ measurements is considered and the allowed parameter space for the parameters involved is quantified. In section ~\ref{sec:res}, we present our result for the triple higgs coupling $hhh$ in several scenario and show the consistency of our scan with  the bound from XENON1T experiment   and conclude in Sec.~\ref{sec:conclusion}.

\section{The canonical Inert Doublet Model}
\label{sec:idm-model}
\subsection{Overview}\label{subsec:overview}
The IDM is a slightly extended version of the SM that preserves its heritage in fermion and gauge bosons sectors at tree level. Thus, an additional doublet $\Phi_{2}$ without a {VEV}  was incorporated into the SM Higgs sector, and considering the fact that general $\mathbb{Z}_2$-invariance is imposed, the particles in the inert doublet $\Phi_{2}$ are odd while the remaining fields are even under $\mathbb{Z}_2$. The physical parametrization of the scalar doublets has the form
\begin{align}
    H_{1}=\begin{pmatrix}
              G^{+} \\
              \frac{1}{\sqrt{2}}\left(v+h+iG\right)
          \end{pmatrix}\text{ and }H_{2}=\begin{pmatrix}
                                             H^{+} \\
                                             \frac{1}{\sqrt{2}}\left(S+iA\right)
                                         \end{pmatrix},
    \label{Parametrization}
\end{align}
The most general (dimension 4) $SU(2)_L \times U(1)_Y$ gauge invariant scalar potential with an exact $\mathbb{Z}_2$ symmetry  takes the following form:
\begin{align}
    V & =\mu_{11}^{2}H_{1}^{\dagger }H_{1}+\mu_{22}^{2}H_{2}^{\dagger
    }H_{2}+\eta_{1}\left(H_{1}^{\dagger }H_{1}\right)^{2}+\eta_{2}\left( H_{2}^{\dagger }H_{2}\right)^{2} \notag                            \\
      & +\eta_{3}\left( H_{1}^{\dagger }H_{1}\right)
    \left( H_{2}^{\dagger }H_{2}\right)+\eta_{4}\left( H_{1}^{\dagger }H_{2}\right) \left( H_{2}^{\dagger }H_{1}\right) \notag              \\
      & + \frac{1}{2}\eta_{5}\left[\left( H_{1}^{\dagger }H_{2}\right) ^{2}+\left( H_{2}^{\dagger }H_{1}\right) ^{2}\right],  \label{PotV1}
\end{align}
where $\mu_{11}^{2}$ and $\mu_{22}^{2}$ are mass terms and $\eta_{1,2,3,4,5}$ are quartic couplings.
Electroweak symmetry is broken when $H_1$ gets its VEV  $\langle H_{1}\rangle_{0}= v$ GeV while $H_2$
stays with a vanishing VEV. $G^\pm$ and $G$ are the charged and neutral Goldstone bosons respectively,
which are absorbed by the longitudinal component of $W^\pm$ and $Z$ to acquire their masses. The Higgs spectrum of the model contains: $i)$ the SM Higgs boson $h$ and a neutral scalars boson $S$ which are defined as scalars transforming to CP symmetry in a even way, $ii)$ a pseudo-scalar field, $A$, changing odd under CP symmetry and $iii)$ the fields $H^{\pm}$ that are the charged scalar bosons.

The scalar bosons masses are given by:
\begin{eqnarray}
    && m_{h}^{2} = 2 \eta_{1}v^{2} = -2 \mu_{11}^2,  \label{Ei1} \\
    && m_{S}^{2} = \mu_{22}^2+\eta_L v^{2},  \label{Ei2} \\
    && m_{A}^{2} = \mu_{22}^2+\eta_S v^{2},  \label{Ei3} \\
    && m_{H^{\pm }}^{2} =\mu_{22}^2+\frac{1}{2}\eta_{3}v^{2},  \label{Ei4}
\end{eqnarray}
where the new expressions $\eta_{L,S}$ are as follows
\begin{equation}
    \eta_{L} = \frac{1}{2}\left(\eta_{3}+\eta_{4}+\eta_{5}\right)\,\,,\,\,\eta_{S} = \frac{1}{2}\left(\eta_{3}+\eta_{4}-\eta_{5}\right).
    \label{Ei5}
\end{equation}
Moreover, the splitting among the neutral, charged scalar masses as well as the $\mu_{22}^2$ might be expressed by
\begin{eqnarray}
    && \Delta m_0^2 = m_{S}^{2} - m_{A}^{2} = \eta_{5} v^2,  \label{Ei6} \\
    && \Delta m_1^2 = m_{S}^{2} + m_{A}^{2} - 2 m_{H^{\pm }}^{2}  = \eta_{4} v^2,  \label{Ei7} \\
    && \Delta m_2^2 = m_{H^{\pm }}^{2} - \mu_{22}^2  = \frac{1}{2} \eta_{3} v^2.  \label{Ei8}
\end{eqnarray}
which could be worthwhile manner to give viable values for the particular quartic couplings $\eta_3$, $\eta_4$ and $\eta_5$. As it can be seen from above, $\eta_5$ can be determined from the splitting between the square of $S$ and $A$ masses. The IDM Higgs sector is thus described by the following six parameters, which we choose to be
\begin{eqnarray}
    \mathcal{P}=\{\mu_{22}^2,\,\eta_2,\,m_h,\,m_S,\,m_A,\,m_{H^\pm}\}\label{Ei9}
\end{eqnarray}
in which the scalar field $h$ fully mimics the SM Higgs boson in mass and couplings with fermionic and gauge bosonic fields. For the self-Higgs coupling $hhh$, as for its coupling to the charged scalar bosons, both can be derived at the tree-level from the scalar potential in Eq.(\ref{PotV1}). They read,
\begin{eqnarray}
    && \eta_{hhh} = -3 m_h^2/v,  \label{Ei10} \\
    && \eta_{hH^{\pm}H^{\mp}}= \frac{2}{v} \left( \mu_{22}^2 - m_{H^{\pm }}^{2} \right)=-\eta_3 v \label{Ei11}
\end{eqnarray}

\noindent
Here, as clearly indicated by Eq.(\ref{Ei11}), the coupling $hH^{\pm}H^{\mp}$ is directly related to the quartic coupling $\eta_3$, which is in turn related to the splitting between the charged scalar boson mass square and $\mu_{22}^2$ : $\left( \mu_{22}^2 - m_{H^{\pm }}^{2} \right)$. Furthermore, the value of the leading order self coupling $\eta_{hhh}$ is fixed by the experimental measurement of the Higgs masse $m_h$.

\subsection{Theoretical and Experimental Constraints}
\label{subsec:theo-exp-const}
In the following, we sum up all theoretical constraints that must be imposed on the scalar sector, for the IDM  to be consistent with the principles of electroweak symmetry breaking. Firstly, the unitarity constraints puts bounds on the amplitude of partial waves \cite{Akeroyd:2000wc}, which in turn curtail the values of the coupling constants. The latter go into the composition of the $\mathcal{S}$-scattering matrix eigenvalues given by
\begin{eqnarray}
    && e_{1,2} = \eta_3 \pm \eta_{4} \textrm{ , } e_{3,4} = \eta_3 \pm \eta_{5} \nonumber \\
    && e_{5,6} = \eta_3  + 2 \eta_4 \pm 3 \eta_5 \nonumber \\
    && e_{7,8} = - \eta_1 - \eta_2 \pm \sqrt{(\eta_1 + \eta_2)^2 + \eta_4^2} \nonumber \\
    && e_{9,10} = -3 \eta_1 - 3 \eta_2 \pm \sqrt{9(\eta_1 - \eta_2)^2 + (2 \eta_3 + \eta_4)^2}  \nonumber \\
    && e_{11,12} = - \eta_1 - \eta_2 \pm \sqrt{(\eta_1 - \eta_2)^2 + \eta_5^2}
\end{eqnarray}
which must all be below $8 \pi$. Pursuant to such requirement, a compact constraint on
$\eta_{1,2}$ stands out to be : $\eta_{1,2} \leq 4 \pi/3$. We also recall that the potential is also perturbative, so we will impose that all the quartic couplings in Eq.(\ref{PotV1}) to be $|\eta_i| \leq 8 \pi$.

Secondly, in order to have one minimum value, the scalar potential of the IDM model must be bounded from below in all directions of the space-field when the scalar fields become quite large. This corresponds to :
\begin{eqnarray}
    && \eta_1>0,\,\, \eta_2>0,\,\, \eta_3 + 2
    \sqrt{\eta_1 \eta_2} > 0 \nonumber\\
    &&{\rm and} \,\ \eta_3 + \eta_4 - |\eta_5| > 2 \sqrt{\eta_1 \eta_2}
\end{eqnarray}
Similarly, a sufficient but not necessary condition to get neutral charge-conserving vacuum, should be imposed to the potential \cite{Ginzburg:2010wa}:
\begin{equation}
    \eta_{4}\leq|\eta_{5}|
\end{equation}
while the following constraints
\begin{eqnarray}
    m_h^2, m_{S}^2, m_{A}^2, m_{H^\pm}^2 > 0 \quad \textrm{and}  \quad v^2 \sqrt{\eta_1 \eta_2} + \mu_{22}^2 > 0
\end{eqnarray}
are vital to having an inert vacuum \cite{Ginzburg:2010wa}.

Thirdly, the quantum corrections parameterized by the oblique parameters S, T and U \cite{Peskin:1991sw}, make it possible to scrutinize the NP in the electroweak domain, and, accordingly, their effects on the $W$ and $Z$ bosons self energies may also restrain the IDM space parameter (see Ref.\cite{Barbieri:2006dq} for the analytic $S$ and $T$ formulas in the IDM). Those parameters are mainly sensitive to the above splitting between the scalar states, and when fixing $\Delta U$ at zero, their allowable values, according to the latest global fit of electroweak precision data read  \cite{ParticleDataGroup:2022pth}:
\begin{equation}
    \Delta S = -0.01\pm0.07,\, \Delta T = 0.04 \pm 0.06
\end{equation}
Overall, throughout our study, such oblique parameters are performed at $2\sigma$ using the PDG results, and other collider constraints \cite{Belyaev:2016lok,Cao:2007rm,Dercks:2018wch,Ilnicka:2015jba,ATLAS:2017fak,CMS:2018yfx} that satisfy lower bonds on the new scalar bosons masses have been considered.

\noindent
Experimentally, constraints from direct searches at LEP, adapted from the production neutralinos and charginos in the framework of the Minimal Supersymetric Standard Model, bound the scalars masses as follows~\cite{B_langer_2015,Lundstr_m_2009} :
\begin{eqnarray}
    && m_{H^\pm} > 80\ \text{GeV},\ \text{max}(m_A, m_S)> 100\ \text{GeV}, \\
    && m_A+m_S > m_Z\ \text{ and } m_A + m_{H^\pm} > m_W
\end{eqnarray}
Additionally, to examine more widely the IDM space parameter, we evaluate the signal strength, defined as,
\begin{eqnarray}
    \mu_{V\gamma} = \frac{ \sigma(pp \to h) } { \sigma^{SM}(pp \to h) } \times \frac{Br(h \to V\gamma)}{Br(h_{\textrm{SM}} \to V \gamma)},\,\, \text{with}\,\,V=\gamma \, \text{or}\,\, Z, \label{eq:muVgatheo}
\end{eqnarray}
and compare it to the experiment measurement values. Signal strengths for both decay modes have been analyzed at 13 TeV center of mass energy. The best fit values for the $h \to \gamma \gamma$ decay mode, as reported by ATLAS \cite{ATLAS:2022tnm} and CMS \cite{CMS:2021kom}, are respectively given by:
\begin{eqnarray}
\mu_{\gamma\gamma}^{\rm ATLAS} &=& 1.04^{+0.10}_{-0.09}, \label{mugagaATLAS}\\
\mu_{\gamma\gamma}^{\rm CMS}  &=&1.12^{+0.09}_{-0.09} \label{mugagaCMS}.
\end{eqnarray}
For the $h \to \gamma Z$ decay mode, CMS \cite{CMS:2022ahq} reports a best fit value of:
\begin{equation}
\mu_{Z\gamma}^{\rm CMS}=2.4\pm0.9. \label{muZgaCMS}
\end{equation}
Moreover, with the upcoming HL-LHC project, increasing sensitivity is expected to boost the corresponding forward-looking measurements, potentially achieving high accuracy of \cite{Cepeda:2019klc}:
\begin{equation}
    \label{eq:muZVgaexp-hl-lhc}
    \mu_{\gamma\gamma}^{\text{HL-LHC}}=1 \pm 0.04\,\,\, \text{and} \,\,\, \mu_{Z\gamma}^{\text{HL-LHC}}=1 \pm 0.23.
\end{equation}

However, after extensive work, ATLAS and CMS combined their data and found the first proof of the $h\to Z\gamma$ decay \cite{CMS:2023csd}, with a statistical significance of 3.4 standard deviation. The observed signal strength is: $\mu_{Z\gamma}^{exp}=2.2\pm 0.7$
which is consistent with the SM theoretical expectation within 1.9 standard deviations.\\
Note that in the IDM, the signal strength of $h \to \gamma\gamma$ and $h\to Z\gamma$ reduces to the ratio of the corresponding Branching ratio normalized to the SM value. This is mostly caused by the fact that, at the leading order,  $\sigma(gg \to h)$ is the same in both the IDM and SM.

The above LHC constraints on the signal strength $\mu_{\gamma\gamma}^{exp}$ are $\mu_{Z\gamma}^{exp}$ together with the invisible decay constraints are the only ones to consider since the Higgs is  SM like, all Higgs production cross sections as well as Higgs decay are exactly as in the SM.

\section{$h\to \gamma\gamma$ and $h\to Z\gamma$ in the IDM}
\label{sec:signal-strength}
At one loop level, the di-photon Higgs decay (together with photon+Z boson decay scheme) could be mediated by the newly charged scalar boson, $H^\pm$  in addition to the SM contribution dominated by the W loops. The corresponding decay widths are given explicitly by \cite{Gunion:1989we,Djouadi:2005gi,Djouadi:2005gj,Krawczyk:2013pea,Fortes:2014dia}, and take the form
\begin{align}
    \Gamma(h \to \gamma\gamma)=     & \frac{G_F\,\alpha^2\,m_h^3}{128\sqrt{2}\pi^3}\bigg|\sum_{f}\underbrace{Q_f^2 N_c A^{\gamma \gamma}_{\frac{1}{2}}(\tau_f)}_{C_{top}}+ \underbrace{A^{\gamma \gamma}_{1}(\tau_W)}_{C_W}  \underbrace{-\frac{m_W}{g\,m_{H^\pm}^2}\,\eta_{hH^{\pm}H^{\mp}}\,A_{0}^{\gamma \gamma}(\tau_{H^\pm})}_{C_{H^\pm}}\bigg|^2
    \label{Ei12}                                                                                                                                                                                                                                                                                                                                       \\
    \Gamma (h \rightarrow Z\gamma)= & \frac{G_F^2\,\alpha\,m_h^3\,m_W^2}{64\pi^4} \left( 1-\frac{m_Z^2}{m_h^2}\right) \bigg|\sum_f \underbrace{\frac{N_c Q_f \hat v_f}{c_w}A^{\gamma Z}_{\frac{1}{2}}(\tau_f, \lambda_f)}_{C_{top}}+ \underbrace{A_1^{\gamma Z}(\tau_W, \lambda_W)}_{C_W}  \nonumber                                                   \\
    +                               & \underbrace{\frac{v v_{H^{\pm}} }{2m_{H^{\pm}}}\,\eta_{hH^{\pm}H^{\mp}}\,A_0^{\gamma Z}(\tau_{H^{\pm}})}_{C_{H^\pm}}  \bigg|^2 \label{Ei13}
\end{align}
with $ \hat v_f= 2 I_f^3 -4Q_f s^2_W $ and $ v_{H ^{\pm}}=\frac{2c_W^2-1}{c_W}$. Here $ N_c $ is the color factor $N_c=3(1)$ for quarks(leptons) and  $Q_f$ stands for the electric charge  of a particle  in the loop.  $ \alpha $ is  the fine-structure constant. The $A^{VV}_{1/2}$, with $VV\equiv\gamma\gamma,\gamma Z$, (for the fermions, $f=t,b,\tau$) as well as the $A^{VV}_{1}$  (for the W boson) and $A^{VV}_{0}$ (for the charged scalars, $ H^{\pm}$ ) are three dimensionless form factors for spin-$1/2$, spin-$1$ and spin-$0$ particles, which can be expressed using the Passarino-Veltamn functions \cite{Passarino:1978jh}. Furthermore, considering Eqs.(\ref{Ei4})-(\ref{Ei11}), it is obvious that $m_{H^\pm}$, $\mu_{22}$ or $\eta_3$ are the relevant keys that could diminish or enhance the IDM prediction of $\Gamma(h \to \gamma\gamma)$ and $\Gamma(h \to Z\gamma)$ compared to the SM one. Since the charged Higgs contribution is proportional to $hH^+H^-$ coupling which is in turn proportional to $\eta_3$, depending on the sign of $\eta_3$ the charged Higgs contribution could be either constructive or destructive with the dominant W loop contributions. Furthermore, it is crucial to highlight that two-loop corrections to the Higgs di-photon decay have been carried out within the IDM, allowing reliable comparisons between theoretical predictions and experimental results. For more details, we refer the reader to Ref.\cite{Aiko:2023nqj}.

\noindent
The model parameters in Eq.(\ref{Ei9}) are randomly scanned within the following ranges:
\begin{eqnarray}
    \label{Ei14}
    && m_h = 125.09\,\text{GeV} \\
    && \eta_2 \in [0,\,4\pi/3]  \ \ \ , \ \ \ \mu_{22}^2 \in [-10^5,\,10^5]\quad (\text{GeV}^2) \nonumber\\
    && m_{S,A}\in [50,\,500]\quad (\text{GeV}) \ \ , \ \ m_{H^\pm} \in [80,\,500]\quad (\text{GeV}) \nonumber
\end{eqnarray}

\begin{figure}[!h]
    \centering
    \includegraphics[width=0.45\textwidth]{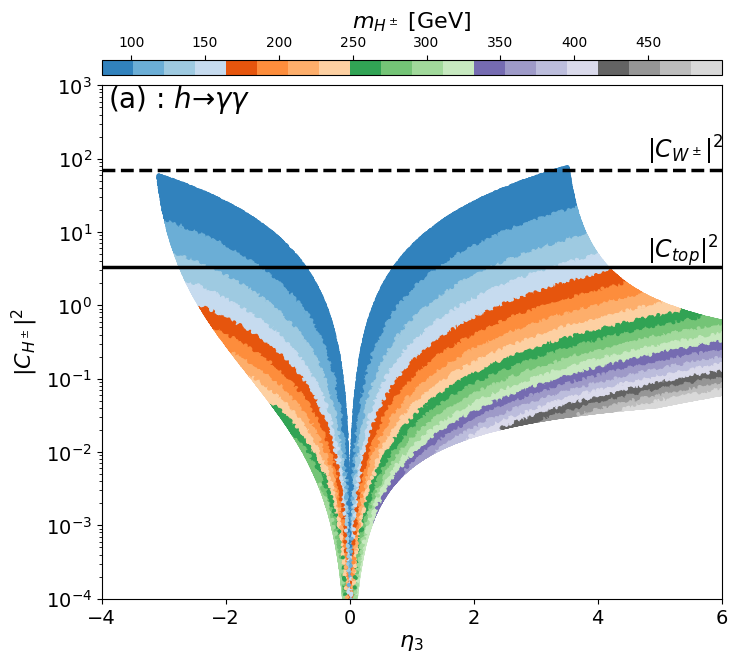}
    \includegraphics[width=0.45\textwidth]{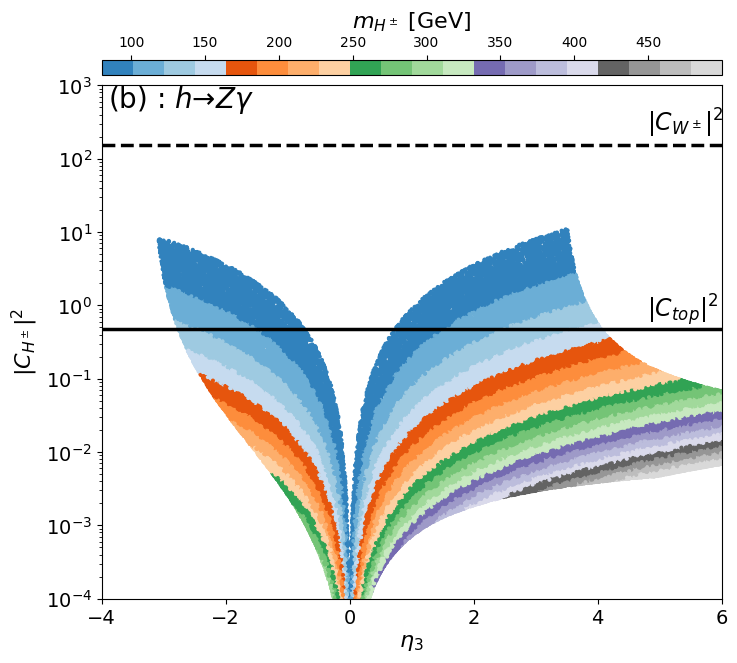}
    \caption{The squared moduli of charged scalar boson contribution $C_{H^\pm}$ for $h \to \gamma\gamma$ (left) and $h \to Z\gamma$ (right)  amplitudes, as a function of $\eta_3$ coupling. The top and $W^\pm$ contributions are shown for comparaison. The color coding exhibits the charged scalar boson mass $m_{H^\pm}$. None of the theoretical or experimental constraints are applied.}
    \label{fig:fig1}
    \centering
\end{figure}

\noindent
In Figure.\ref{fig:fig1}, we illustrate a scatter plot comparing the squared moduli of the various contributions in Eqs.(\ref{Ei12}-\ref{Ei13}) without regard for any theoretical constraints on the scalar potential parameters (for the fermionic one, we have drawn only the top quark contribution).
As it can be observed, the diagrams mediated by internal vectorial bosons ($W^\pm$)
have the most expressive contribution compared to the top quark, either for $h \to \gamma\gamma$ (Figure.\ref{fig:fig1}-(a)) or $h \to Z\gamma$ (Figure.\ref{fig:fig1}-(b)). Nevertheless, the charged scalar boson contribution is non-negligible, and could  contributes significantly by some few orders of magnitude. It is obvious that for small $\eta_3\approx 0$,
$|C_{H\pm}|^2$ is very suppressed. While for large value of $\eta_3$ together with light charged scalar boson mass in the range $[80,150]$ GeV one can see that $|C_{H^\pm}|^2$ could be between $|C_{top}|^2$ and $|C_{W^\pm}|^2$, and can get to $|C_{W^\pm}|^2$ value for the $\gamma\gamma$ mode as it can be seen form Figure.\ref{fig:fig1}-(a). For heavier charged scalar boson $\geq 150$ GeV, its contribution drops below $|C_{top}|^2$. However, such 
contribution could be either constructive or destructive with the W loops.
\begin{figure}[t]
    \centering
    \includegraphics[width=0.45\textwidth]{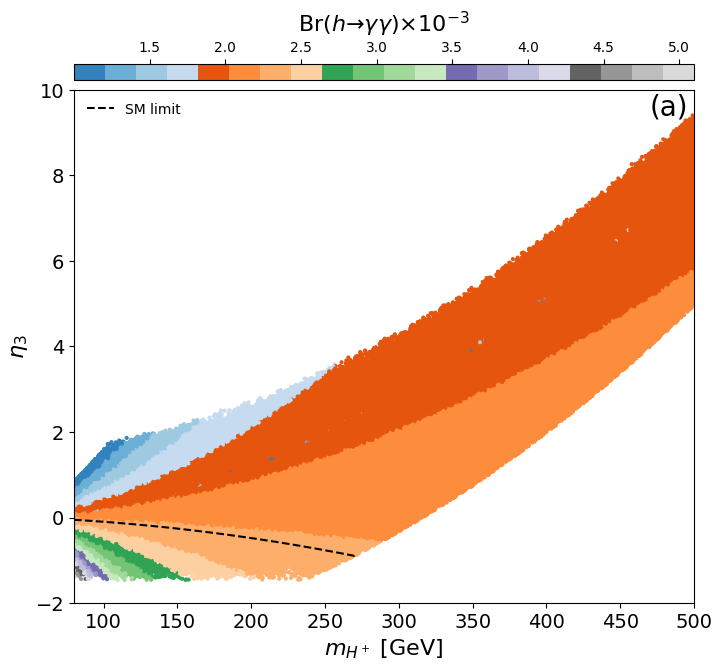}
    \includegraphics[width=0.45\textwidth]{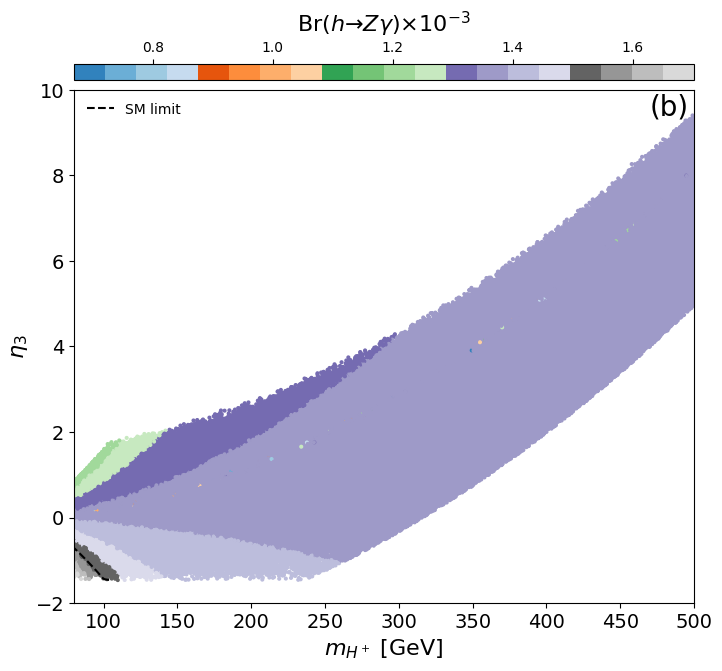}
    \caption{Branching ratios: $\text{Br}(h\to\gamma\gamma)$ (left panel) and $\text{Br}(h \to Z\gamma)$ (right panel) as a function of $m_{H^\pm}$ and the coupling $\eta_3$. Theoretical constraints as well as LEP requirements are applied.}
    \label{fig:fig2}
    \centering
\end{figure}

Figure.\ref{fig:fig2} could promote the aforementioned interference. Indeed, by considering the LEP constraints, we illustrate $\text{Br}(h\to \gamma \gamma)$ (left) and $\text{Br}(h\to Z \gamma)$ (right) as a function of $m_{H^\pm}$ and $\eta_3$. For $\text{Br}(h\to \gamma \gamma)$, it is clear that there is a large area of parameter space where the branching ratio is smaller than the SM value and this is because of the destructive interference between the $W^\pm$ and $H^\pm$ loops. One can also see a small region with negative $\eta_3$ and rather light charged scalar boson $m_{H\pm}$ where  $\text{Br}(h\to \gamma \gamma)$ is greater than the SM value, and this is due to the constructive interference between $W^\pm$ and $H^\pm$ loops.\\
For the $Z\gamma$ decay mode, it can be observed that, except for a very small region with $\eta_3<0$ and charged scalar boson masses less than 100 GeV, the $\text{Br}(h\to Z \gamma)$ is below the SM value on the whole. This region corresponds to constructive interference between the SM Higgs  and the charged scalar loops. Conversely, in the opposite case, there is a large area in the charged scalar boson mass and $\eta_3$ parameter space where the $\text{Br}(h\to Z \gamma)$ is smaller than the SM value.

\section{Results and Discussion}
\label{sec:res}
In this section, we illustrate the result of our scan both for  the triple coupling $hhh$ as well as the consistency of our result with  XENON1T  experiment on the spin-independent dark-matter-nucleon scattering cross-section. To address that purpose, we explore some numerical consequences distinguishing between two cases:
\begin{enumerate}
    \item [$\bullet$] Degenerate case where $m_S=m_A=m_{H^\pm}=m_{\Phi}$, in other words $\Delta m_0=\Delta m_1=0$, that will enable us to avoid electroweak precision observables (EMPO) constraints in IDM.
    \item [$\bullet$] Quasi-degenerate case where $m_S \ne m_A = m_{H^\pm}$ and $m_S \le m_h/2$.
\end{enumerate}
and whose only those that obey the boundary parameters allowed by the limitations of the previous theoretical as well as the experimental constraints are survived.

\subsection{Radiative Corrections to $hhh$}
\label{subsec:radiahhh}
In this section, we will illustrate the impact of $\mu_{\gamma\gamma}^{exp}$ measurement on the trilinear coupling $\eta_{hhh}$ within the IDM. The latter has been the subject of several studies BSM such as: MSSM \cite{Hollik:2001px}, 2HDM \cite{Kanemura:2004mg} and the IDM \cite{Arhrib:2015hoa}. They have applied renormalization techniques to this issue and show its sensitivity to the NP effects BSM. Here, with the aim of addressing the IDM towards such measurements, we revisit the additional contribution to the following process,
\begin{equation}
    \label{Ei15}
    h(q) \to h(k_1) + h(k_2)
\end{equation}
\begin{figure}[!h]
    \includegraphics[width=0.45\textwidth]{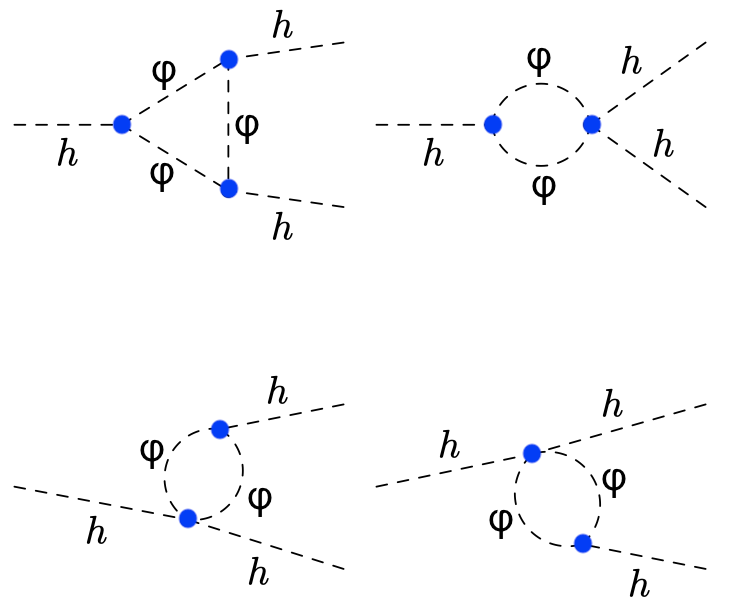}
    \caption{Diagrams contributing, at one-loop, to  the radiative corrections of the trilinear coupling $\eta_{hhh}$ in the IDM. The $\varphi$ stands for any scalar $S$, $A$ and $H^\pm$.}
    \label{fig:fig3}
    \centering
\end{figure}
where $q$ (resp. $k_1$ and $k_2$) denotes the 4-momenta of the incoming particle satisfying off shell condition $q^2\neq m_h^2$ (resp. the 4-momenta of the outgoing particles satisfying on shell condition $k_1^2=k_2^2=m_h^2$), is calculated from the Feynman diagrams depicted in Figure.\ref{fig:fig3}

As is also being demonstrated in \cite{Arhrib:2015hoa}, the contribution of the IDM is purely bosonic backed by the $h_{125}$ Higgs boson couplings to the new inert scalars. Hence, the corresponding amplitude, is given by
\begin{eqnarray}
    \Gamma_{hhh}^{loop} (q^2,m_\Phi^2)&&=\frac{\eta_3^{2}m_{W}s_{W}}{8e\pi^2}
    \biggl( B_{0}(q^2,m_{H^\pm}^{2},m_{H^\pm}^{2})+2B_{0}(m_{h}^{2},m_{H^\pm}^{2},m_{H^\pm}) \nonumber \\
    &&+\frac{2 \eta_3 m_{W}^2s_{W}^2}{ \pi \alpha } C_{0}(q^2,m_{h}^{2},m_{h}^{2},m_{H^\pm}^{2},m_{H^\pm}^{2},m_{H^\pm}^{2})\biggr) \nonumber \\
    &&+\frac{(\eta_3+\eta_4+\eta_5)^{2} m_{W}s_{W}}{16 e \pi^2}\biggl( B_{0}(q^2,m_{S}^{2},m_{S}^{2})+2B_{0}(m_{h}^{2},m_{S}^{2},m_{S}^{2}) \nonumber\\
    && \times\frac{2(\eta_3+\eta_4+\eta_5) m_{W}^2s_{W}^2}{\pi \alpha}C_{0}(q^2,m_{h}^{2},m_{h}^{2},m_{S}^{2},m_{S}^{2},m_{S}^{2})\biggr) \nonumber \\
    &&+\frac{(\eta_3+\eta_4-\eta_5)^{2} m_{W}s_{W}}{16e\pi^2}\biggl(B_{0}(q^2,m_{A}^{2},m_{A}^{2})+2B_{0}(m_{h}^{2},m_{A}^{2},m_{A}^{2}) \nonumber \\
    &&\times\frac{2(\eta_3+\eta_4-\eta_5) m_{W}^2s_{W}^2}{\pi \alpha}C_{0}(q^2,m_{h}^{2},m_{h}^{2},m_{A}^{2},m_{A}^{2},m_{A}^{2})\biggr)
    \label{Ei16}
\end{eqnarray}
\begin{figure}[!h]
    \includegraphics[width=0.45\textwidth]{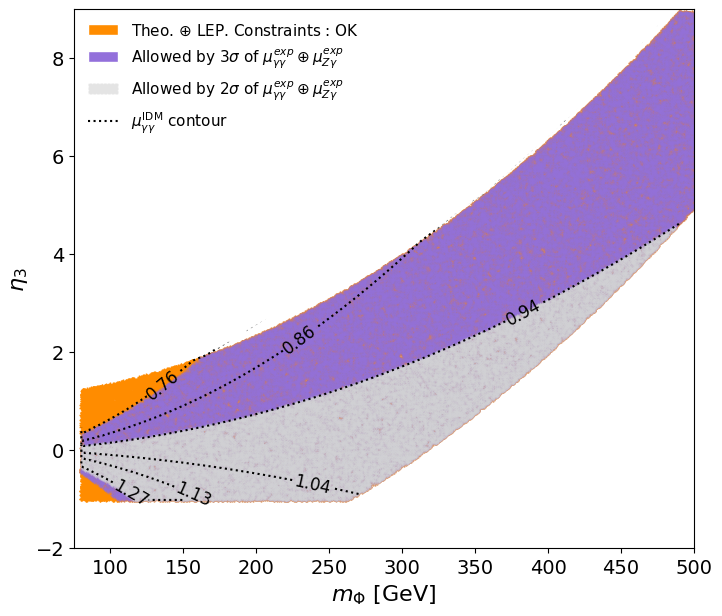}
    \includegraphics[width=0.45\textwidth]{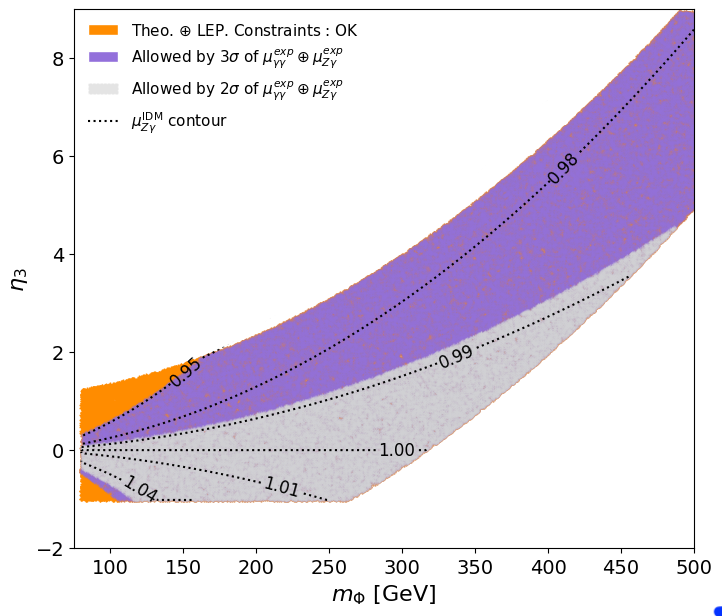}\\
    \includegraphics[width=0.45\textwidth]{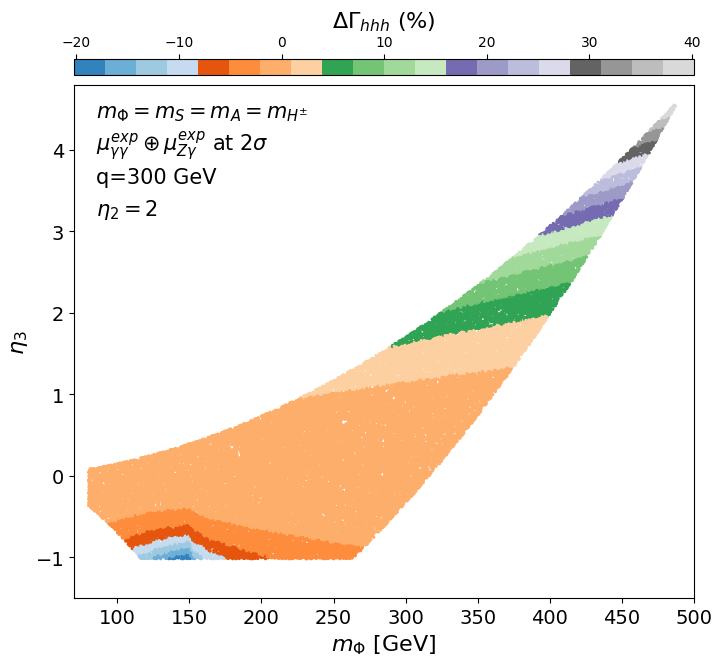}
    \caption{Upper panel: correlation between $m_{\phi}=m_S=m_A=m_{H^\pm}$ and $\eta_3$ with the dot-dashed line corresponds to the $\mu_{\gamma\gamma}^{\rm IDM}$ (left) and $\mu_{Z\gamma}^{\rm IDM}$ (right)  contours. Lower panel: the
    $\Delta\Gamma_{hhh}$ variation within the $(\eta_3-m_{\Phi})$ plane at $95\%$ .C.L. We set $\eta_3=2$.}
    \label{fig:fig4}
    \centering
\end{figure}
with $B_0$ and $C_0$ are the Passarino-Veltman functions \cite{Passarino:1978jh} is not UV finite, it was important to add the corresponding counter-term, $\delta\Gamma_{hhh}^{loop}$, and evaluate them by calculating the necessary and sufficient renormalization constants. For more details, we refer the reader to \cite{Arhrib:2015hoa}.

\noindent
In line with our purpose in this study, we redefine a ratio that involves the previous quantities as,
\begin{equation}
    \label{Ei17}
    \Delta\Gamma_{hhh}=\frac{\Gamma_{hhh}^{loop}+\delta\Gamma_{hhh}^{loop}-\Gamma_{hhh}^{tree}}{\Gamma_{hhh}^{tree}}
\end{equation}
where the coupling $\Gamma_{hhh}^{tree}=-3m_h^2/v$ represents the trilinear coupling at tree level.

By considering the degenerate case where $100\,\text{GeV} \le m_{\Phi} \le 500\,\text{GeV}$, a random scan over the IDM parameter space is performed taking into account the effects of the ATLAS and CMS experimental measurement constraints on $\mu_{\gamma\gamma,Z\gamma}$ at 95$\%$ C.L.. The upper panel of Figure.\ref{fig:fig4} shows the allowed parameter space in the $(\eta_3-m_{\Phi})$ plane with $\eta_2=2$. Thus, in addition to all the theoretical requirements, it further seem to point to the fact that at $99\%$ C.L. of $\mu_{\gamma\gamma,Z\gamma}^{exp}$ measurements; the permissible space parameter requires $m_{\Phi} \lesssim 500$ GeV and $\eta_3 \lesssim 9$ as reflected by the gray region. Nevertheless, scaling down to $\pm2\sigma$ provides a reduced surface area (orange) with an upper limit can extend to $488\,\text{GeV}$, but still ruling out any enhancement for the $\text{Br}(h \to Z\gamma)/\text{SM}$ and $\text{Br}(h \to \gamma\gamma)/\text{SM}$ above 4.5$\%$ and 27$\%$, respectively. However, it should further be emphasized that such experimental measurements require a positive value of $\mu_{22}^2$ in order to validate the values in Eqs.(\ref{mugagaATLAS}-\ref{mugagaCMS}-\ref{muZgaCMS}).

The lower panel of Figure.\ref{fig:fig4} addresses only the consistent points with $\mu_{Z\gamma,\gamma\gamma}^{exp}$ at $2\sigma$ C.L., and depicts the relative corrections to the triple coupling $hhh$ in the $(\eta_3-m_{\Phi})$ plane, with the assumption that $q=300$ GeV. At first sight, one can see how the space shrank so drastically compared to \cite{Arhrib:2015hoa,Falaki:2023tyd} results, and the enhancement they found for the radiative corrections went down dozens of times, and had narrowed to only $-20\%$ for low $m_{\Phi}$ and $36\%$ for $m_{\Phi}$ around 485 GeV. Furthermore, it is clear that the $|\Delta\Gamma_{hhh}|$ goes up steadily over $\eta_3$ for fixed value of $m_{\Phi}$, and could reach its maximum around  $\eta_3=-1$ (around $\mu_{22}^2\approx50000\,\text{GeV}^2$) for $m_{\Phi}\approx150$ GeV. Such latter point is amplified by the the threshold channel $h^\ast\to \Phi\Phi$ that is open, where the momentum of the off shell Higgs boson is $q=2m_\Phi=300$ GeV.

\subsection{Invisible Higgs decay and $\Delta\Gamma_{hhh}$}
\label{subsec:radiahhh}
A further advantage of IDM enables to explain the invisible Higgs sector as it embeds $S$ or $A$, both of which  can be considered as stable WIMP particles, and then the SM Higgs can invisibly decay into $h\to SS$ or $h\to AA$ which can modify the total width of the Higgs boson and have significant impact on the LHC results. There exist several studies on the limit of the invisible decay of the SM Higgs boson, conducted by both ATLAS and CMS \cite{ATLAS:2020kdi,CMS:2022qva}. The most recent limit, as reported by ATLAS in \cite{ATLAS:2023tkt}, reads:
\begin{equation}
    \label{Ei18}
    \text{Br}_{\text{inv}} < 11\% \,\, \text{at} \,\, 95\% \,\, \text{C.L.}
\end{equation}
and will be applied. The partial width for the process in Eq.(\ref{Ei18}) is expressed by,
\begin{equation}
    \label{Ei19}
    \Gamma(h \to SS, AA) = \frac{v^2\,\eta_{L,S}^2}{8\,\pi\,m_h}.\sqrt{1-\frac{4\,m_{S,A}^2}{m_h^2}}
\end{equation}
and is proportional to $\eta_{L,S}^2$ given in Eq.(\ref{Ei5}). \\
Subsequently, as one of our aims through this study is to shed light on how far $\mu_{Z\gamma,\gamma\gamma}^{exp}$ together with other Higgs observable issues and the radiative corrections $\Delta\Gamma_{hhh}$ can examine the hidden sector within the IDM if the inert particle $S$ is light enough, we have slightly changed our set of parameters to be
\begin{eqnarray}
    \label{Ei19}
    && m_h = 125.09\,\gev, \hspace*{0.25cm} m_S \in [10,\,62.5]\,\gev, \hspace*{0.25cm} m_{A},\,m_{H^\pm}  \in [80,\,500]\,\gev, \nonumber\\
    && \eta_2 = 2\,, \hspace*{0.25cm} \mu_{22}^2 \in [-4\times10^4,\,10^5]\,\gev^2
\end{eqnarray}

\begin{figure}[!ht]
    \centering
    \includegraphics[width=0.45\textwidth]{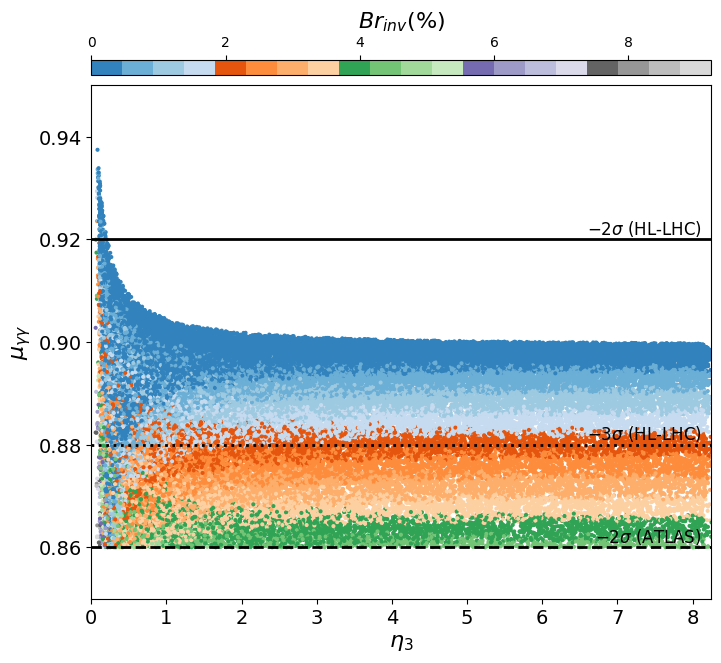}
    \includegraphics[width=0.45\textwidth]{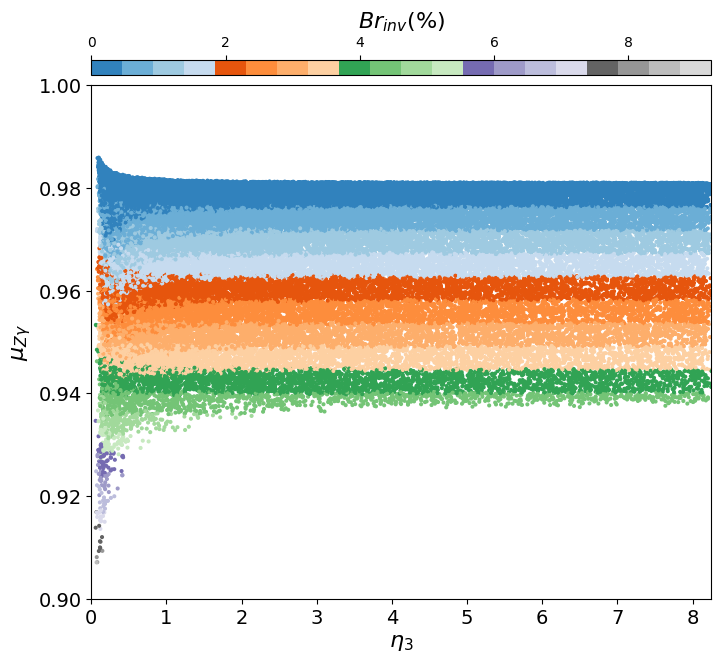}
    \caption{The $\mu_{\gamma\gamma}$ (left) and $\mu_{Z\gamma}$ (right) as a function of $\eta_3$ in the quasi-degenerate case, with the color coding shows the invisible branching ratio. For both plots $\eta_3$ is fixed to be 2 and $10\,\gev \le m_S \le 62.5\,\gev$. The solid, dotted and dashed black lines in the left panel correspond to the more conservative limit from LHC-HL at $-2\sigma$, $-3\sigma$ and ATLAS at $-2\sigma$, respectively.}
    \label{fig:fig5}
\end{figure}

In Figure.\ref{fig:fig5}, we show results for $\mu_{\gamma\gamma}$ (left) and $\mu_{Z\gamma}$ (right)~as a function of $\eta_3$. The invisible decay result is given by the color coding, while the solid, dotted and dashed black lines in the left indicate the expected values at $2\sigma$ and $3\sigma$ on the $\mu_{\gamma\gamma}$ from HL-LHC~\cite{Cepeda:2019klc} and ATLAS collaboration \cite{ATLAS:2022tnm}. For $\mu_{Z\gamma}$, we note that our findings remain above the $-2\sigma=0.6\,(0.54)$ CMS (HL-LHC) values (which are outside the plots) and close to the unity. At first glance, we can observe that the BSM deviation approaches a plateau as $\eta_3$ increases either for $\mu_{\gamma\gamma}$ or $\mu_{Z\gamma}$. This behavior can be explained by the destructive interference between the charged scalar and SM contribution. Regarding the branching ratio $Br_{inv}$, it remains mostly below the aforementioned recent limit if we consider either the $\mu_{Z\gamma}^{\rm CMS}$ or $\mu_{\gamma\gamma}^{\rm ATLAS}$ at 95$\%$ C.L.. However, the situation is drastically modified, and the $Br_{inv}$ does not exceed one $1\%$ if the $\mu_{\gamma\gamma}^{\rm HL-LHC}$ at 95$\%$ C.L. is considered. Such branching ratio get suppressed for small values of $\eta_3 \sim 0$ (or $m_{H^\pm}\sim 80$ GeV).

\begin{figure}[!hb]
    \includegraphics[width=0.45\textwidth]{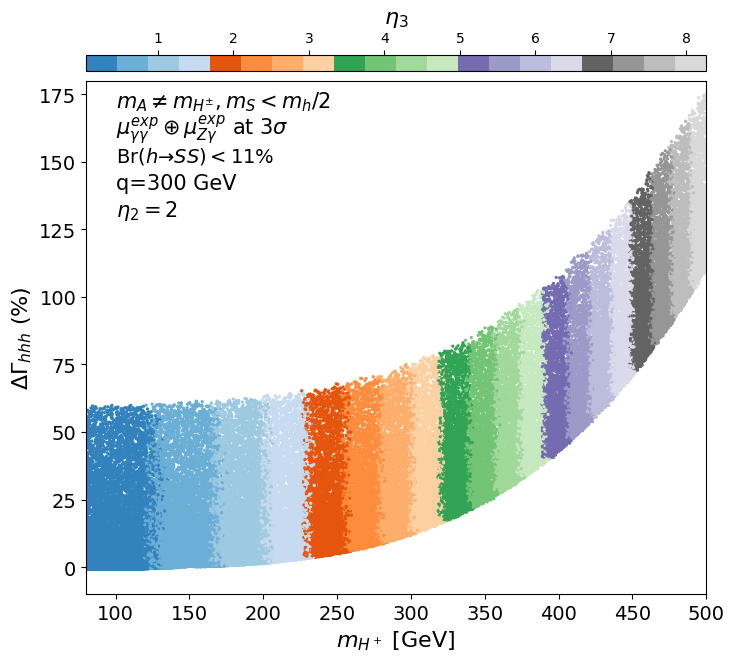}
    \includegraphics[width=0.45\textwidth]{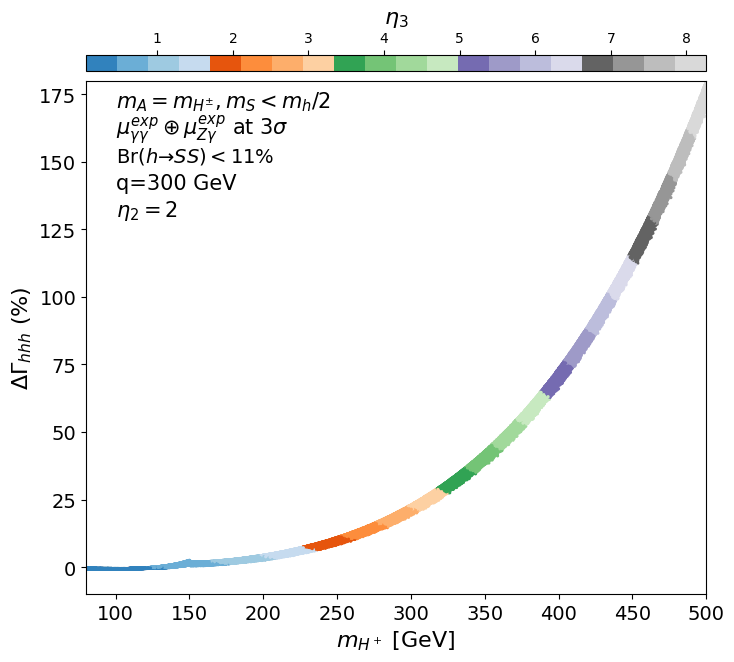}\\
    \caption{Variation of the $\Delta\Gamma_{hhh}$ in the $(m_{H^\pm},\eta_3)$ plane for both cases : $80\,\gev \le m_A,\,m_{H^\pm} \le 500\,\gev$ (left) and $80\,\gev \le m_A = m_{H^\pm} \le 500\,\gev$ (right). We set for both : $m_{S} < m_{h}/2$, $q=300$ GeV and $\eta_2=2$ .}
    \label{fig:fig6}
    \centering
\end{figure}

\noindent
In this regard, Figure.\ref{fig:fig6} exhibits two possible connection between radiative corrections and the trilinear coupling $\eta_3$ in the presence of an invisible decay mode of $h_{125}$. A sizeable rate of the radiative corrections shows its ongoing reliance on the charged scalar bosons loops. Nevertheless, it should be emphasized that at $2\sigma$ of $\mu_{\gamma\gamma}^{exp}$, the Higgs invisible decay scenario is completely ruled out and no significant evidence can be proven. But as a byproduct of our analysis, stretching towards $3\sigma$ provides a likelihood, even the slightest. So, based on the relevant foregoing analysis, it is clearly seen from the left panel in Figure.\ref{fig:fig6} that radiative corrections could not exceed $63\%$ at 95$\%$ C.L. at the HL-LHC, expecting an upper limit on the quartic coupling $\eta_3 \lesssim 0.35$. To achieve this, a light charged scalar boson, i.e $m_{H^\pm}\lesssim 100$ GeV, is needed. Nevertheless, such correction becomes almost entirely suppressed if the $A$ and $H^\pm$ are degenerated as can be drawn from right side in Figure.\ref{fig:fig6}.

\subsection{Dark matter search}
\label{subsec:DMsearch}
\begin{figure}[!h]
    \includegraphics[width=0.45\textwidth]{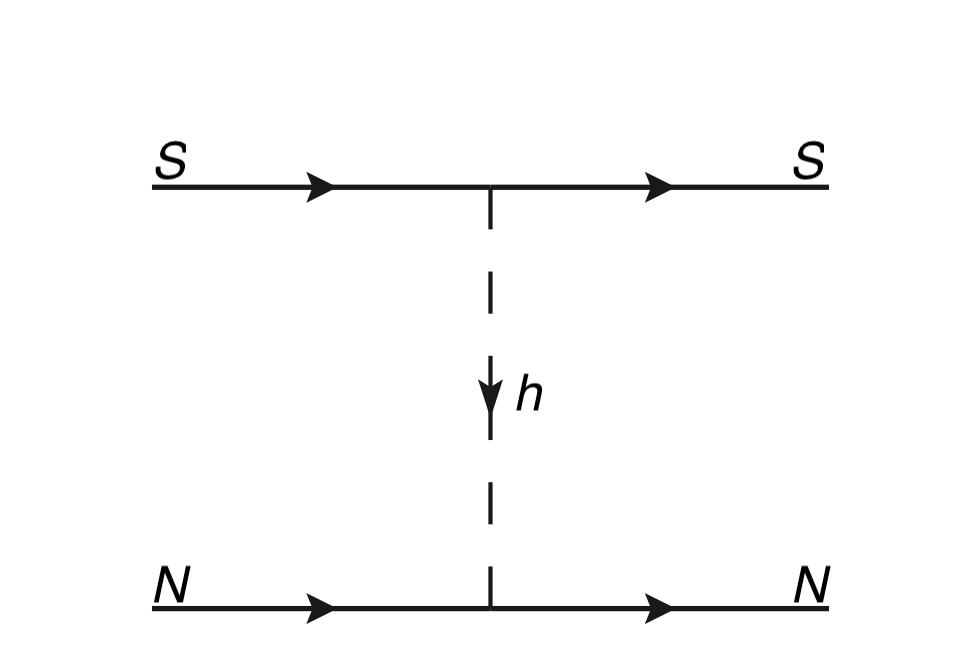}
    \caption{Feynman diagram for the elastic spin-independent scattering cross section of Dark Matter with nucleon mediated by the SM-like Higgs $h$ in the IDM}
    \label{fig:fig7}
    \centering
\end{figure}

In this section, we consider the implication of $\mu_{\gamma\gamma}^{exp}$ measurement on the dark matter within the IDM. Such mysterious stuff that fills the universe may be detected either indirectly through looking for the products of dark matter interactions, especially the SM ones namely bosons, quarks and leptons \cite{Feng:2010gw,Hooper:2009zm}, or directly via interaction with ordinary matter by \cite{Bae:2022dgj}. In our study, we focus on the latter and assume that the inert 
scalar boson $S$, is considered as good candidate to address this fundamental issue. The Feynman diagram describing the scattering between $S$ and nucleon mediated by the observed $h=h_{125}$ is given by Figure.\ref{fig:fig7}
\begin{figure}[b]
    \includegraphics[width=0.45\textwidth]{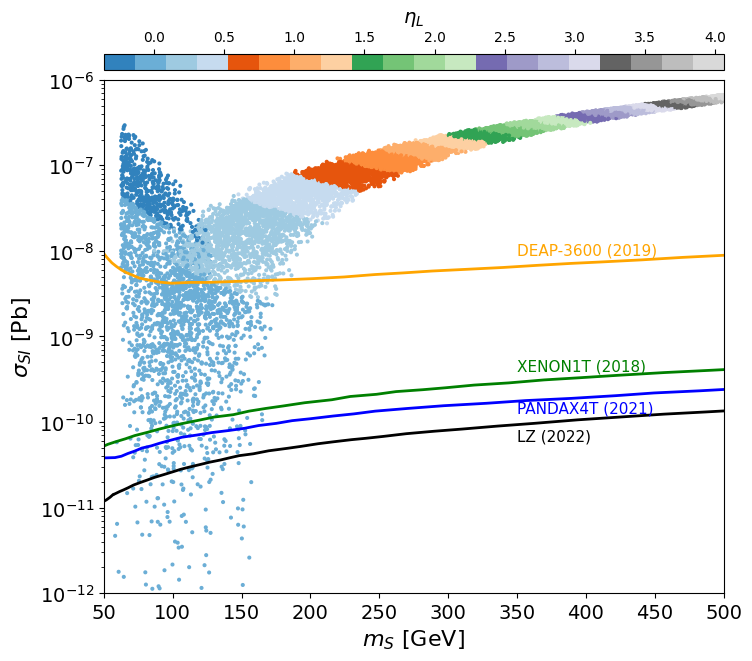}
    \caption{Variation of the spin-independent cross section $\sigma_{SI}$ with dark matter mass $m_S$ where the color coding shows the variation of the $\eta_L$ coupling. The DEAP-3600 \cite{DEAP-3600:2017uua}, XENON1T \cite{XENON:2018voc}, PandaX-4T \cite{PandaX-4T:2021bab},  and LZ \cite{LZ:2022ufs} resulting $90\%$ C.L. upper limits are shown. Our parameters have been set in the non-degenerate case with $\lambda_2=2$.}
    \label{fig:fig8}
    \centering
\end{figure}

Additionally, the spin-independent scattering cross-section relevant to this process can be expressed by
\begin{equation}
    \label{Ei20}
    \sigma_{SI} = \frac{\eta_L^2\,f^2}{4\,\pi}\frac{\mu^2 m_N^2}{m_h^4\,m_S^2}
\end{equation}
where $f=0.32$ \cite{Giedt:2009mr}, $m_N$ and $\mu=m_N m_S/(m_N+m_S)$ denote respectively the Higgs-nucleon coupling, the nucleon mass and the reduced mass of dark matter and nucleon.

Figure.\ref{fig:fig8} delimits the achievable points of space parameter for the Higgs-portal DM particle, $S$ in the IDM.
In generating this plot, further constraints are applied, in addition to those detailed in section \ref{subsec:theo-exp-const}, by imposing an upper limit on the DM relic density, $\Omega_{DM}h^2 \le 0.12 $ \cite{ParticleDataGroup:2020ssz}, calculated using the micrOMEGAs 5.2 framework \cite{Belanger:2020gnr}. We have also imposed: $\lambda_2=2$ as for previous plots and $\text{Br}_{\text{inv}} < 11\%$ while the inert scalar mass has been varied (in GeV) in the range $[50,\,500]$. The first thing to note is that Higgs–DM mass whether light (i.e. $M_S \leq 60$ GeV, which requires $|\eta_L|\le0.02$) or wedged in $[60,180]$ (GeV), could be probed by the Xenon1T \cite{XENON:2018voc} experimental sensitivity bands, and its distribution fundamentally shifted in the PandaX-4T experiment and LZ \cite{LZ:2022ufs}. However, there is still a narrow regions that are ruled out by those present experiments e.g.

\begin{itemize}
    \item [$i)$] the window with $m_h/2\le m_S/\text{GeV} \le 135$ with $\mu_{22}^2 \le 2.5\,\text{GeV}^2$ and $\eta_L \le 0$.
    \item [$ii)$] Also, the region where $m_S/\text{GeV} \ge 175$ together with $\eta_L \ge 0.25$,
\end{itemize}
but they are still under experimental scrutiny and might be tested in future experiences \cite{DARWIN:2016hyl} widening an enough space parameters.

\section{Conclusion}
\label{sec:conclusion}
We have revisited in this paper the IDM as extension BSM by examining the two decay modes $\gamma\gamma$ and $Z\gamma$. In line with the actual data on the corresponding signal strengths $\mu_{\gamma\gamma}^{exp}$ and $\mu_{Z\gamma}^{exp}$, a particular focus is made on the implications of such measurements over many  physical observables.
Our results have shown that $\mu_{\gamma\gamma}^{exp}$ may clamp down the allowed space parameter for the IDM. Notably, at 2$\sigma$ of $\mu_{\gamma\gamma}^{exp}$, the possibility of a Higgs invisible decay scenario is unequivocally excluded, leaving no substantial evidence to support it. This result highlights the stringent limits that current experimental data impose on alternative Higgs decay channels.
As we have demonstrated, the dark matter had a foothold in this paper, and it is highly sensitive to the $\mu_{\gamma\gamma}^{exp}$ measurement. The corresponding cross section has been evaluated, and besides posing a challenge to detect the DM in the universe, it may also provide a way to constrain some model parameters.

\bibliography{citerefs}
\end{document}